\documentclass[aps]{revtex4}
\usepackage{amssymb,amsmath,epsfig}
\usepackage[colorlinks=true, pdfstartview=FitV, linkcolor=blue, citecolor=red, urlcolor=magenta, breaklinks=true]{hyperref}
\begin{document}
\title{Quantum correction to the entropy of noncommutative BTZ black hole}
\author{M. A. Anacleto$\,^{1}$}
\email{anacleto@df.ufcg.edu.br}
\author{F. A. Brito$\,^{1,2}$}
\email{fabrito@df.ufcg.edu.br}
\author{A. G. Cavalcanti$\,^{1}$}
\email{arthurlovb@hotmail.com}
\author{E. Passos$\,^{1}$}
\email{passos@df.ufcg.edu.br}
\affiliation{$\,^{1}$Departamento de F\'{\i}sica, Universidade Federal de Campina Grande
Caixa Postal 10071, 58429-900 Campina Grande, Para\'{\i}ba, Brazil\\
$\,^{2}$ Departamento de F\'isica, Universidade Federal da Para\'iba, Caixa Postal 5008, 58051-970 Jo\~ao Pessoa, Para\'iba, Brazil}
\author{J. Spinelly}
\email{jspinelly@uepb.edu.br}
\affiliation{Departamento de F\'isica-CCT, Universidade Estadual da Para\'iba
\\
Juv\^encio Arruda S/N, Campina Grande, PB, Brazil}

\begin{abstract} 
In this paper we consider the generalized uncertainty principle (GUP) in the
tunneling formalism via Hamilton-Jacobi method to determine the quantum-corrected Hawking temperature and entropy for noncommutative BTZ black hole. In our results  we obtain several types of corrections including the expected logarithmic correction to the area entropy associated with the  noncommutative BTZ black holes. 
We also show that the area entropy product of the noncommutative BTZ black holes is dependent on mass and by analyzing the nature of the specific heat capacity we have observed that the noncommutative BTZ black hole is stable at some range of parameters.

\end{abstract}
\maketitle
\pretolerance10000
\section{Introduction}
In recent years a number of studies in three-dimensional gravity has increased due to the discovery of various aspects
of black-hole solutions. 
Black holes constitute an important class of exact solutions of Einstein's equations which are characterized by mass
($M$), electric charge ($Q$) and angular momentum ($J$)~\cite{Frolov} and play a central role in both classical and quantum gravitational physics.
In the literature there are various ways to implement the noncommutative in the physics of black holes.
Thus, interest in the study of noncommutative black holes have been investigated by many authors in recent years (for a review see \cite{Nicolini:2008aj}). 
In particular, the noncommutative {Ba\~nados}-Teitelboim-Zanelli (BTZ) black holes were first analyzed in~\cite{Banados} 
and in~\cite{Chang} the noncommutative BTZ metric was found based on the three dimensional equivalence between gravity and the Chern-Simons theory   which is a 3-dimensional topological quantum field theory and using the Seiberg-Witten map with the commutative BTZ solution~\cite{BTZ}.  
{The BTZ black hole is a solution of (2+1) dimensional gravity with negative cosmological
constant and has become an important field of investigations~\cite{Eune:2013qs}.  }
It is now well accepted that three-dimensional gravity is an excellent laboratory in order to explore and test some of the ideas behind the AdS/CFT correspondence~\cite{ADS}.
In~\cite{Anacleto:2014cga} the authors analyzed the gravitational Aharonov-Bohm effect due to BTZ black hole
in a noncommutative background.
Moreover, in~\cite{Jafar} has been analyzed the behavior of a particle test in the noncommutative BTZ space-time.
{The thermodynamic properties of the charged BTZ black hole were investigated in~\cite{Hendi:2015wxa,Shi-Wei}. }

A semiclassical approach considering the Hawking radiation as a tunneling phenomenon across the horizon has been proposed in recent years \cite{Parikh, ec.vagenas-plb559}.
In this approach the positive energy particle created just inside the horizon can tunnel through the geometric barrier quantum mechanically, and it is observed as the Hawking flux at infinity.
There are several approaches to obtain the Hawking radiation and the entropy of black holes. One of them is the 
Hamilton-Jacobi method which is based on the work of Padmanabhan and collaborators~\cite{SP} and also the effects of the self-gravitation of the particle are discarded.
In this way, the method uses the WKB approximation in the tunneling formalism for the computation of the imaginary part of the action. The authors Parikh and Wilczek~\cite{Parikh} using the method of radial null geodesic determined the Hawking temperature and in~\cite{Jiang} this method was used by the authors for calculating the Hawking temperature for different spacetimes. In Ref.~\cite{Banerjee} has been analyzed Hawking radiation considering  self-gravitation and back reaction effects in tunneling formalism. It has also been investigated in~\cite{Silva12} the back reaction effects for self-dual black hole using the tunneling formalism by Hamilton-Jacobi method. 
In~\cite{Majumder:afa} has been studied the effects of the generalized uncertainty principle (GUP) in the tunneling formalism for Hawking radiation to evaluate the quantum-corrected Hawking temperature and entropy of a Schwarzschild black hole.
Moreover, the authors in~\cite{Becar:2010zza} have discussed the Hawking radiation for acoustic black hole using  tunneling formalism 
and in~\cite{Anacleto:2015mma} the thermodynamical properties of self-dual black holes, using the Hamilton-Jacobi version of the tunneling formalism were investigated.
It was analyzed in~\cite{Faizal:2014tea} the corrections for the thermodynamics of black holes
assuming that the GUP corrected entropy-area relation is universal for all black objects.

In the literature there are several works on the statistical origin of black hole entropy  --- see for instance ~\cite{Wilczek, Magan:2014dwa,Solodukhin:2011gn,mhorizon}. 
In Ref.~\cite{Kaul}, Kaul and Majumdar computed the lowest order
corrections to the Bekenstein-Hawking entropy. They find that the leading correction is logarithmic.
On the other hand, in Ref.~\cite{Carlip:2000nv} was shown that there is an additional logarithmic corrections that depend on conserved charges.
In addition, for an understanding of the origin of black hole entropy, 
the brick-wall method proposed by  't Hooft has been used for calculations on black holes. Thus, according to  't Hooft, black hole entropy is just the entropy of quantum fields outside the black hole horizon. However, when one calculates the black hole statistical entropy by this method, to avoid the divergence of states density near black hole horizon, an ultraviolet cut-off must be introduced. 
In Ref.~\cite{Rinaldi}  was investigated (1 + 1)-dimensional acoustic black hole entropy by the brick-wall method. 

The other related idea in order to cure the divergences is to consider models in which the Heisenberg uncertainty relation is modified.
Thus, using the modified Heisenberg uncertainty relation the divergence in the brick-wall model are eliminated as discussed in~\cite{Brustein}.
The statistical entropy of various black holes has also been calculated via corrected state density of the GUP~\cite{XLi}. 
Thus, the results show that near the horizon quantum state density and its statistical entropy are finite. In~\cite{KN} a relation for the corrected states density by GUP has been proposed.
The authors in~\cite{ABPS} using a new equation of state density due to GUP~\cite{Zhao}, the statistical entropy of a 2+1-dimensional rotating acoustic black hole has been analyzed. It was shown that considering the effect due to GUP on the equation of state density,  no cut-off is needed \cite{Zhang} and the divergence in the brick-wall model disappears.

In this paper, inspired by all of these previous work we shall focus on the Hamilton-Jacobi method to determine the entropy of a noncommutative BTZ black hole using the GUP and considering the WKB approximation in the tunneling formalism to calculate the imaginary part of the action in order to determine the Hawking temperature and entropy for  BTZ black holes. We anticipate that we have obtained the Bekenstein-Hawking entropy of BTZ black holes and its quantum corrections that are logarithm and also of other types. 

The paper is organized as follows. In Sec.~\ref{NCBTZ} we briefly review the noncommutative BTZ black holes geometry and address the entropy product. We notice that differently of the commutative case,  for noncommutative BTZ black holes such a product depends on the mass of the black holes. 
We also use the Hamilton-Jacobi method to determine the Hawking temperature and that the noncommutative correction for Hawking temperature occurs only at second order in the noncommutativity parameter. In Sec.~\ref{GUP-sec} we consider the GUP in the tunneling formalism via the Hamilton-Jacobi method to find the quantum corrections to the Hawking temperature, entropy and specific heat capacity of a noncommutative BTZ black hole. Finally in Sec.~\ref{conclu} we present our final
comments.

\section{Noncommutative BTZ black holes}\label{NCBTZ}
In this section we consider the metric of the BTZ black hole
in a noncommutative background given by~\cite{Chang,Anacleto:2014cga}:
\begin{eqnarray}
ds^{2} = - Fdt^{2} + N^{-1}dr^{2} +2r^{2}N^{\phi}dtd\phi+ \left(r^{2} - \frac{\theta B}{2}\right)d\phi ^{2},
\label{metrica BTZ}
\end{eqnarray}
where the metric components are
\begin{eqnarray}
&&F = \frac{r^{2} - r^{2}_{+} - r^{2}_{-}}{l^{2}} - \frac{\theta B}{2l^2},
\label{F}
\\
&&N = \frac{1}{r^{2}l^{2}}\left[(r^{2} - r^{2}_{+})(r^{2} - r^{2}_{-}) 
- \frac{\theta B}{2}(2r^{2} - r^{2}_{+} - r^{2}_{-})\right],
\label{N}
\\
&& N^{\phi}=-\frac{r_+r_-}{l r^2}.
\end{eqnarray}
{Here $B$ is the magnitude of a $U(1)$ flux in a noncommutative $U (1, 1)\times U (1, 1)$ Chern-Simons theory and $\theta$ is the noncommutative parameter of dimension $length^2$. The Seiberg-Witten map is carried out up to first order in $\theta$ --- see Refs.~\cite{Chang} for further details.}
For the noncommutative  BTZ black hole, the event horizons are given by
\begin{eqnarray}
\hat{r}_{\pm}^2=r_{\pm}^2+\frac{\theta B}{2} + {\cal O}(\theta^2),
\end{eqnarray}
and 
\begin{eqnarray}
r^2_{\pm}=\frac{l^2M}{2}\left[1\pm\sqrt{1-\left(\frac{J}{Ml}\right)^2}  \right],
\label{Raio comutativo}
\end{eqnarray}
where $r_{+}$  is the outer  event horizon and $r_{-}$ is the inner event horizon of the commutative BTZ black hole.
Note that for $ \theta=0 $ the event horizons of the commutative case are recovered.

In order to analyze the product of entropy,   we will first consider the product and sum of the horizon radii that are given by

\begin{eqnarray}
\hat{r}_{+}\hat{r}_{-}=\sqrt{ r^2_{+}r^2_{-}+\frac{\theta B}{2}(r^2_{+} +r^2_{-})+O(\theta^2)}
=\sqrt{\frac{l^2J^2}{4}+\frac{\theta B l^2M}{2}+O(\theta^2)}=\frac{lJ}{2}\left[1+\frac{\theta B M}{J^2} + O(\theta^2)\right],
\end{eqnarray}
and 

\begin{eqnarray}
\hat{r}^2_{+}+\hat{r}^2_{-}={r}^2_{+}+{r}^2_{-}+\theta B=l^2M+\theta B +O(\theta^2).
\end{eqnarray}
Observe that the product and the sum depend on mass parameter. On the other hand, for $ \theta=0 $, the product 
$ \hat{r}_{+}\hat{r}_{-}=lJ/2$ is independent on mass.

Let us consider that $\tilde{S}_{\pm} = 4\pi\hat{r}_{\pm}$ is the entropy of the noncommutative  BTZ black holes. 
Thus, the product  $\tilde{S}_{+}\tilde{S}_{-}$ is given by
\begin{eqnarray}
\tilde{S}_{+}\tilde{S}_{-} &=& 16\pi ^{2} \hat{r}_{+}\hat{r}_{-} 
=16\pi ^{2} \sqrt{\frac{l^2 J^2}{4} + \frac{\theta Bl^2M}{2} +O(\theta^2)},
\nonumber\\
&=& 8\pi^{2} l J  \left[ 1 + \frac{\theta B M}{J^2}\right] + O(\theta ^2).
\end{eqnarray}
Notice that, {at least up to first order in $\theta$}, the entropy product of the noncommutative BTZ black holes is dependent on mass. 
On the other hand, for $ \theta=0 $, the result is independent on mass~\cite{Pradhan:2015ela}.
It is conjectured that the product of the areas for multi-horizon stationary black holes are
in some cases independent on the mass of the black hole~\cite{Ansorg}.
However, there are studies in the literature where the areas product is dependent on the mass~\cite{Visser}. 
For example, it has been shown that for Schwarzschild-de Sitter black hole in (3+1) dimensions the product of event horizon area and cosmological horizon area is {\it not} mass independent.
Recently, it was also shown in Ref.~\cite{Anacleto:2013esa} for acoustic black hole that the universal aspects of the areas product depends only on quantized quantities such as analogues of conserved electric charge and angular momentum.

The metric of noncommutative BTZ black hole can be rewritten as
	\begin{eqnarray}
	ds^{2} = -fdt^{2} + Q^{-1}dr^{2}-\frac{J}{r}rd\phi dt + \left(1-\frac{\theta B}{2r^{2}}\right)r^{2}d\phi ^{2},
	\label{metrica BTZ1}
	\end{eqnarray}
where
\begin{eqnarray}
	&&f = -M + \frac{r^{2}}{l^{2}} - \frac{\theta B}{2l^{2}},
	\label{f}
\\
	&&Q = -M + \frac{r^{2}}{l^{2}} +\frac{J^2}{4r^2}- \frac{\theta B}{2}\left(\frac{2}{l^{2}} - \frac{M}{r^{2}}\right).
	\label{Q}
	\end{eqnarray}
At this point we will consider the case where $J = 0$. Thus, near the event horizon of a noncommutative BTZ black hole, we can rewrite the metric~(\ref{metrica BTZ1}) as follows
	\begin{eqnarray}
	ds^{2} = -\tilde{f} dt^{2} + \tilde{Q}^{-1}dr^{2} + \left(1-\frac{\theta 		B}{2r^{2}}\right)r^{2}d\phi ^{2},
	\label{metrica-BTZ1-aprox}
	\end{eqnarray}
where $ \tilde{f}=f^{\prime}(\hat{r}_{+})(r - \hat{r}_{+}) $ and $ \tilde{Q}=Q'(\hat{r}_{+})(r - \hat{r}_{+}) $.

Now we use the Hamilton-Jacobi method to determine the Hawking temperature. 
Using the Klein-Gordon equation for a scalar field $\Phi$ given by
\begin{eqnarray}
	\left[\frac{1}{\sqrt{-g}}\partial _{\mu}(\sqrt{-g}g^{\mu \nu}\partial _{\nu}) - \frac{m^{2}}{\hbar ^{2}}\right] \Phi = 0 ,
	\label{Klein Gordon}
\end{eqnarray}
and applying the WKB approximation
	\begin{eqnarray}
	\Phi = \exp\left[\frac{i}{\hbar}I(t,r,x^{i})\right],
	\label{WKB}
	\end{eqnarray}
we obtain	
	\begin{eqnarray}
	g^{\mu\nu}\partial _{\mu}I\partial _{\nu}I + m^{2} = 0,
	\label{KG-WKB}
	\end{eqnarray}
that in terms of the metric (\ref{metrica-BTZ1-aprox}), becomes
	\begin{eqnarray}
	-\frac{1}{\tilde{f}}(\partial _{t}I)^{2} + \tilde{Q}(\partial _{r}I)^{2} + \frac{1}{r^{2}}(\partial _{\phi}I)^{2} + m^{2} = 0.
	\label{KG-WKB componente}
	\end{eqnarray}
Now we can assume a solution to the Klein-Gordon equation via separation of variables as follows
	\begin{eqnarray}
	I = -Et + W(r) + J_{\phi}\phi,
	\label{I}
	\end{eqnarray}
where $J_{\phi}$ is a constant. 
By substituting (\ref{I}) into equation (\ref{KG-WKB componente})
and solving for $W(r)$ the spatial part of the classical action reads
\begin{eqnarray}
	W(r) =\int _{C} \frac{\sqrt{E^2 - f'(\hat{r}_{+})(r - \hat{r}_{+}) \left(\frac{2J_{\phi}^2}{2r^2 - \theta B} 
	+ m^2\right)}}{\sqrt{f'(\hat{r}_{+})Q'(\hat{r}_{+})}}
	=\frac{2\pi i}{\kappa}E,
	\label{W}
\end{eqnarray}
where
	\begin{eqnarray}
	\kappa = \sqrt{f'(\hat{r}_{+})Q'(\hat{r}_{+})} = \sqrt{\frac{4\hat{r}_{+}^2}{l^4} - \frac{2\theta B M}{l^2\hat{r}_{+}^2}}.
	\label{k}
	\end{eqnarray}
On the other hand, the probability of a particle overcoming the potential barrier is given by		
	\begin{eqnarray}
	\Gamma = \exp[-2{\rm\, Im}(I)] \quad  \Rightarrow  \quad   \Gamma = \exp\left(-\frac{4\pi E}{\kappa}\right).
	\label{Gamma}
	\end{eqnarray}
Now comparing (\ref{Gamma}) with the Boltzmann factor
	$\exp({-{E}/{\tilde{T}_{H}}})$
we obtain the Hawking temperature of the BTZ black hole in a noncommutative background
\begin{eqnarray}
\label{ThB}
	{\tilde T}_{H} = \frac{\kappa}{4\pi}=\frac{1}{4\pi}\sqrt{\frac{4\hat{r}_{+}^2}{l^4} - \frac{2\theta B M}{l^2\hat{r}_{+}^2}}
	=\frac{\hat{r}_+}{2\pi l^2}\left(1- \frac{\theta B r_+^2}{4\hat{r}_{+}^4} \right)+\cdots,
	\label{TH}
\end{eqnarray}
that in terms of $r_+=\sqrt{l^2M}$, we have	
\begin{eqnarray}
	{\tilde T}_H=\frac{{r}_+}{2\pi l^2}\left(1- \frac{\theta^2 B^2 }{16{r}_{+}^4} \right)+\cdots
	=T_{h}-\frac{\theta^2 B^2}{256\pi^4l^8 T^3_{h}}+\cdots,
	\end{eqnarray}	
where $ T_h={{r}_+}/{(2\pi l^2)}$ is the Hawking temperature of the BTZ black hole. The above result shows that the non-commutative correction for Hawking temperature occurs only at second order in the parameter $ \theta $.	


{Now for the case of $ J\neq 0 $ the line element of Eq. (\ref{metrica BTZ1}) can be written as follows	
\begin{eqnarray}
\label{mdiag}
	ds^{2} = -{\cal F}dt^{2} + {\cal Q}^{-1}dr^{2}+ \left(1-\frac{\theta B}{2r^{2}}\right)r^{2}d\varphi ^{2},
	\end{eqnarray}
where
\begin{eqnarray}
	&&{\cal F} = -M + \frac{r^{2}}{l^{2}} +\frac{J^2}{4r^2}+\frac{\theta B J^2}{8r^4}- \frac{\theta B}{2l^{2}},
	\label{f1}
\\
	&&{\cal Q} = -M + \frac{r^{2}}{l^{2}} +\frac{J^2}{4r^2}- \frac{\theta B}{2}\left(\frac{2}{l^{2}} - \frac{M}{r^{2}}\right),
	\label{Q1}
\\
&& d\varphi=d\phi- \frac{J}{2 \left(1-{\theta B}/{2r^{2}}\right)r^{2}}dt.
\label{tc}
\end{eqnarray}
{Here we have performed the coordinate transformation (\ref{tc}), in order to write the metric (\ref{metrica BTZ1}) in the diagonal form --- Eq. (\ref{mdiag}). Thus, we can apply the same procedure done previously to determine the Hawking temperature.
In this case, near the event horizon of the BTZ black hole we have the tunneling probability $ \Gamma=\exp[-4\pi E/\bar{\kappa}] $, where $ \bar{\kappa}=\sqrt{{\cal F}^{\prime}(\hat{r}_{+}){\cal Q}^{\prime}(\hat{r}_{+})} $ is the surface gravity and so comparing $ \Gamma $ with the Boltzmann factor $\exp({-{E}/{{\cal T}_{H}}})$ the Hawking temperature is given by }
\begin{eqnarray}
	{\cal T}_{H} &=&\frac{\bar{\kappa}}{4\pi}= \frac{\sqrt{{\cal F}^{\prime}(\hat{r}_{+}){\cal Q}^{\prime}(\hat{r}_{+})}}{4\pi}
	\nonumber\\
	&=&\frac{2\hat{r}_{+}}{4\pi l^2}\left( 1- \frac{l^2J^2}{4\hat{r}^4_{+} }\right)
	\sqrt{1 -\left[ \frac{\theta B l^2 M}{2\hat{r}_{+}^4} +\frac{\theta Bl^2 J^2}{4\hat{r}^6_{+} } 
	\left( 1-\frac{l^2 M }{2\hat{r}^2_{+} } -\frac{l^2 J^2 }{4\hat{r}^4_{+} } \right)\right] \left( 1- \frac{l^2J^2}{4\hat{r}^4_{+} }\right)^{-2}  + {\cal O}(\theta^2) }.
\end{eqnarray}
At the limit $ \theta\rightarrow 0 $ we obtain the temperature of the commutative BTZ black hole
\begin{eqnarray}
\label{ThJ}
	{\cal T}_{h} &=& \frac{{r}_{+}}{2\pi l^2}\left( 1- \frac{l^2J^2}{4{r}^4_{+} }\right).
\end{eqnarray}	
Note that by comparing the Eqs. (\ref{ThB}) and (\ref{ThJ}) the quantity $ \theta B l^2M $ in (\ref{ThB}) due to noncommutativity mimics an angular momentum type contribution.

From Eq. (\ref{Q1}) we obtain the mass of the noncommutative black hole that is given by
\begin{eqnarray}
M&=&\left(1-\frac{\theta B}{2\hat{r}^2_+} \right)^{-1}\left[ \frac{\hat{r}_+^{2}}{l^{2}} +\frac{J^2}{4\hat{r}_+^2} 
- \frac{\theta B}{l^2}  \right],
\nonumber\\
&=&\frac{\hat{r}_+^{2}}{l^{2}} +\frac{J^2}{4\hat{r}_+^2} +\theta B\left( \frac{J^2}{8\hat{r}_+^4}-\frac{1}{2l^2}\right) + {\cal O}(\theta^2),
\end{eqnarray}
and since $ M $ is a function of $ \hat{r}_+ $ and $ J $, $ M=M(\hat{r}_+,J) $, we can write
\begin{eqnarray}
\label{dM}
dM=\frac{\partial M}{\partial\hat{r}_+}d\hat{r}_{+}+\Omega dJ,
\end{eqnarray}
where $\Omega =\frac{\partial M}{\partial J} $ and 
\begin{eqnarray}
\label{dm2}
\frac{\partial M}{\partial\hat{r}_+}=\frac{2\hat{r}_+}{l^{2}} -\frac{J^2}{2\hat{r}_+^3}\left(1+\frac{\theta B}{\hat{r}_+^2}\right) + {\cal O}(\theta^2)= \frac{2\hat{r}_+}{l^{2}}\left( 1- \frac{l^2J^2}{4\hat{r}^4_{+} }\right) - \frac{\theta B J^2}{\hat{r}_+^5}  .
\end{eqnarray}

In order to analyze the entropy we consider the first law of the thermodynamics of black holes, thus we have
\begin{eqnarray}
\label{dS}
dM={\cal T}_HdS+\Omega dJ.
\end{eqnarray}
By comparing (\ref{dM}) and (\ref{dS}) we obtain the following relation for the entropy
\begin{eqnarray}
\label{ds2}
dS=\frac{1}{{\cal T}_H}\frac{\partial M}{\partial\hat{r}_+}d\hat{r}_{+}.
\end{eqnarray}
Now expanding $ {\cal T}_H^{-1} $ up to first order in $ \theta $ we have
\begin{eqnarray}
\label{expanT}
{\cal T}_H^{-1} =4\pi\left[\frac{2\hat{r}_{+}}{l^2}\left( 1- \frac{l^2J^2}{4\hat{r}^4_{+} }\right)\right]^{-1}
\left\{1 +\left[ \frac{\theta B l^2 M}{4\hat{r}_{+}^4} +\frac{\theta Bl^2 J^2}{8\hat{r}^6_{+} } 
	\left( 1-\frac{l^2 M }{2\hat{r}^2_{+} } -\frac{l^2 J^2 }{4\hat{r}^4_{+} } \right)\right] \left( 1- \frac{l^2J^2}{4\hat{r}^4_{+} }\right)^{-2}   \right\} + {\cal O}(\theta^2).
\end{eqnarray}
Then, substituting (\ref{dm2}) and (\ref{expanT}) into (\ref{ds2}) we find
\begin{eqnarray}
\hat{S}&=&4\pi\int \left\{1 +\left[ \frac{\theta B}{4\hat{r}_{+}^2} +\frac{\theta Bl^2 J^2}{8\hat{r}^6_{+} } 
	\left( 1-\frac{3l^2 J^2 }{8\hat{r}^4_{+} } \right)\right] \left( 1- \frac{l^2J^2}{4\hat{r}^4_{+} }\right)^{-2} \right.
\nonumber\\	
&-&\left. \frac{\theta B J^2}{\hat{r}_+^5}\left[\frac{2\hat{r}_{+}}{l^2}\left( 1- \frac{l^2J^2}{4\hat{r}^4_{+} }\right)\right]^{-1} + {\cal O}(\theta^2)\right\}d\hat{r}_+,
\end{eqnarray}
that in terms of $ r_+ $, we have
\begin{eqnarray}
\label{entj}
\hat{S}&=&4\pi\int\left(1-\frac{\theta B}{4r^2_+} \right) \left\{1 +\left[ \frac{\theta B}{4{r}_{+}^2} 
+\frac{\theta Bl^2 J^2}{8{r}^6_{+} } 
	\left( 1-\frac{3l^2 J^2 }{8{r}^4_{+} } \right)\right] \left( 1+\frac{l^2J^2}{2{r}^4_{+} }\right) \right.
\nonumber\\	
&-&\left. \frac{\theta B l^2J^2}{2{r}_+^6}\left( 1+\frac{l^2J^2}{4{r}^4_{+} }\right) + {\cal O}(\theta^2)\right\}d{r}_+,
\nonumber\\
&=& 4\pi r_{+} +\frac{\pi B^2 \theta^2}{12r_{+}^3}+ \frac{ \pi B \theta l^2J^2}{5 r^5_+} 
+ \frac{7\pi B \theta l^4J^4}{144 r^9_+} +\frac{3\pi B \theta l^6J^6}{ 416r^{13}_+}+\cdots . 
\end{eqnarray}
In the limit of $ \theta=0 $ in (\ref{entj}), the entropy of the commutative BTZ black hole is recovered, i.e., 
$S=4\pi r_{+} $. For  the case $ J=0 $, the entropy becomes 
\begin{eqnarray}\label{ThJ2}
\hat{S}&=& 4\pi r_{+} +\frac{\pi B^2 \theta^2}{12r_{+}^3}+\cdots .
\end{eqnarray}

}


\section{quantum-corrected Hawking temperature and entropy}\label{GUP-sec}	
In this section we consider the GUP in the tunneling formalism via the Hamilton-Jacobi method to find the quantum corrections to the Hawking temperature, entropy and specific heat capacity of a noncommutative BTZ black hole.
Thus our starting point is the GUP~\cite{ADV, Tawfik:2014zca}, which is an extension of \cite{KMM} given by
\begin{eqnarray}
\label{gup}
\Delta x\Delta p\geq \hbar\left( 1-\frac{\alpha l_p}{\hbar} \Delta p +\frac{\alpha^2 l^2_p}{\hbar^2} (\Delta p)^2 \right),
\end{eqnarray}
where $\alpha$ is a dimensionless positive parameter, $ l_p=\sqrt{\hbar G/c^3}={M_p G}/{c^2}\approx 10^{-35}m$ is the Planck length, $ M_p =\sqrt{\hbar c/G}$ is the Plank mass and $ c $ is the velocity of light. 
Since $G$ is the Newtonian coupling constant,  the correction terms in the uncertainty relation (\ref{gup}) are due to the effects of gravity. {Although this formula is written in four spacetime dimensions it also works in 2+1 dimensions under certain assumptions --- see below. {We can offer a simple demonstration of the quadratic part of Eq.~(\ref{gup}). Thus, let us compute the total position uncertainty \cite{maggi} by considering BTZ black holes, i.e.,}
\begin{eqnarray}
\Delta x=\Delta x_1+\Delta x_2 \simeq \frac{\lambda}{\sin{\phi}}+\frac{l}{2}\frac{8G_{3}\Delta M}{\sqrt{8G_{3}M}}\geq\lambda+\frac{4l G_{3}}{\sqrt{8G_{3}M}}\frac{1}{\lambda}.
\end{eqnarray}
Here $\Delta x_1$ is the usual Heisenberg's position uncertainty and $\Delta x_2=r_+(M+\Delta M)-r_+(M)$ is the additional uncertainty due to the BTZ black hole for $J=0$ and $\Delta M\ll M$. This implies the quadratic GUP
\begin{eqnarray}\label{GUP-BTZ}
\Delta x \Delta p\geq1+\alpha^2l\, l_p (\Delta p)^2, \qquad \alpha^2=\frac{4}{\sqrt{8G_{3}M}}, \qquad \Delta p\sim \frac{1}{\lambda},
\end{eqnarray}
where we have reinstated the Newtonian constant at 2+1 dimensions $G_{3}\propto l_p$. However, for latter convenience making $l_p^2=1$ at (\ref{gup}) or $l\,\l_p=1$ at (\ref{GUP-BTZ}) makes the quadratic parts of these GUPs formally the same. Furthermore, the quadratic part of the GUP is naturally consistent to a noncommutative geometric generalization of position space \cite{KMM}. In addition, the linear part of (\ref{gup}) is also consistent with the noncommutativity of the spacetime  \cite{bastos} and Doubly Special Relativity (DSR) theories \cite{Tawfik:2014zca}.
}

Now the equation (\ref{gup}) can be written as follows.
\begin{eqnarray}
\label{xp}
\Delta p\geq \frac{\hbar(\Delta x +\alpha l_p)}{2\alpha^2 l_p^2}
\left(1- \sqrt{1-\frac{4\alpha^2 l_p^2}{(\Delta x +\alpha l_p)^2}}\right),
\end{eqnarray}
where we have chosen the negative sign. 
{Eq. (\ref{xp}) implies a minimum measurable length, $\Delta x\geq (\Delta x)_{min}\approx \alpha l_p $ and a maximum measurable momentum, $\Delta p \leq (\Delta p)_{max}\approx \hbar/(\alpha l_p)  $.}
Since $ l_p/\Delta x $ is relatively small compared to unity we can expand the equation above in Taylor series 
\begin{eqnarray}
\label{p}
\Delta p\geq \frac{1}{2\Delta x}\left[1-\frac{\alpha}{2\Delta x}+ \frac{\alpha^2}{2(\Delta x)^2}+\cdots    \right].
\end{eqnarray}
 Since we have chosen $ G=c=k_B=1 $, we also have $\hbar=1$ and $ l_p=1 $. 
{For $\alpha = 0$ in Eq. (\ref{p}), the uncertainty principle becomes}
\begin{eqnarray}
\label{eqdp}
\Delta x\Delta p\geq 1,
\end{eqnarray}
{where the factor 2 has been absorbed in $ \Delta x $.  Now using the saturated form of the uncertainty principle given in Eq.~(\ref{eqdp}) we can find a bound on the energy of the black hole }
\begin{eqnarray}
E\Delta x\geq1.
\end{eqnarray}
{This result is obtained by considering the standard dispersion relation $ E^2=p^2+m^2$. Then by assuming
$ p\sim\Delta p \geq 1/\Delta x$, we obtain for massless particles the uncertainty in the energy $E =\Delta p\geq 1/\Delta x$. }
Therefore, we can rewrite equation (\ref{p}) in the form
 \begin{eqnarray}
E_{GUP}\geq E\left[1-\frac{\alpha}{2(\Delta x)}+ \frac{\alpha^2}{2(\Delta x)^2}+\cdots    \right].
\end{eqnarray}
So by using Hamilton-Jacobi method, the tunneling probability of a particle with corrected energy  $E_{GUP}$ becomes
\begin{eqnarray}
\Gamma\simeq \exp[-2{\rm Im} ({\cal I})]=\exp\left[\frac{-4{\pi}E_{GUP}}{a}\right],
\end{eqnarray}
{where $ a $ is the surface gravity}. Again, comparing with the Boltzmann factor ($ e^{-E/T} $), we obtain the noncommutative BTZ black hole temperature 
\begin{eqnarray}
T\leq\tilde{T}_H\left[ 1-\frac{\alpha}{2(\Delta x)}+ \frac{\alpha^2}{2(\Delta x)^2}+\cdots   \right]^{-1}.
\end{eqnarray}
{Since from Eqs.~(\ref{ThB}) and (\ref{ThJ}) we can see that the noncommutativity and the angular momentum $J$ play identical roles into the temperature, thus for the sake of simplicity we shall consider $J=0$ from now on. $\tilde{T}_H$ above is given by Eq.~(\ref{ThB}).}

{In this case, near the event horizon of the BTZ black hole, the uncertainty in the position of a particle is of the order of the horizon radius of the black hole BTZ. } So let us now choose $ \Delta x=2\hat{r}_+ $. Thus, we have the corrected temperature due to the GUP
\begin{eqnarray}
\label{Tgup}
T_{GUP}&\leq&\frac{\hat{r}_+}{2\pi l^2}\left(1- \frac{\theta B r_+^2}{4\hat{r}_{+}^4} +\cdots\right)\left(1 - \frac{\alpha}{4\hat{r}_{+}} + \frac{\alpha ^{2}}{8\hat{r}_{+}^{2}}+\cdots\right)^{-1}
\nonumber\\
&=&\frac{\hat{r}_+}{2\pi l^2}\left(1-\frac{\theta B r_+^2}{4\hat{r}_{+}^4} +\cdots\right)
\left(1 + \frac{\alpha}{4\hat{r}_{+}} - \frac{\alpha ^{2}}{8\hat{r}_{+}^{2}}+\cdots\right),
\end{eqnarray}
that in terms of the $ r_+=l\sqrt{M} $, we have
\begin{eqnarray}
T_{GUP}\leq\frac{{r}_+}{2\pi l^2}\left(1-\frac{\theta^2 B^2}{16{r}_{+}^4} +\cdots\right)
\left[1 + \frac{\alpha}{4{r}_{+}}\left(1-\frac{\theta B}{4r_+^2} +\cdots \right) - \frac{\alpha ^{2}}{8{r}_{+}^{2}}
\left(1-\frac{\theta B}{2r_+^2}+\cdots  \right)+\cdots\right]
\end{eqnarray}
or in terms of the Hawking temperature $ T_h=r_+/(2\pi l^2) $ of the BTZ black hole, we obtain
\begin{eqnarray}
T_{GUP}\leq T_{h}-\frac{\theta^2 B^2}{256\pi^4l^8 T^3_{h}}+\frac{\alpha}{8\pi l^2} 
-\frac{\alpha\theta B}{128\pi^3 l^6 T_h^2} 
-\frac{\alpha^2}{32\pi^2 l^4 T_h}+\frac{\alpha^2\theta B}{256\pi^4 l^8 T_h^3}  +\cdots.
\end{eqnarray}
It is interesting to note that the third term in the above equation is independent on the horizon radius.

In the following we will analyze the quantum corrections due to GUP for energy density, specific heat capacity at constant volume and entropy.
{The corrections to the black hole energy density  can be calculated as follows~\cite{Nozari:2008} }
\begin{eqnarray}
\rho _{GUP} = - \frac{3}{l^2} \int S'(A)A^{-2}dA,
\label{Densidade de Energia}
\end{eqnarray}
where, $S'(A) = \frac{dS}{dA}$. Thus
	\begin{eqnarray}
	\rho _{GUP} = \frac{3}{l^2A} - \frac{3}{2l^2}\frac{\alpha\pi}{A^2} 
	+ \frac{2}{l^2}\frac{\pi ^{2}\alpha ^{2}}{A^3} + \frac{3\alpha\theta B\pi ^{3}}{l^2A^4} 
	+\frac{48}{5l^2}\frac{\theta^2 B^2\pi ^{4}}{A^5}- \frac{48}{5l^2}\frac{\alpha ^{2}\theta B\pi ^{4}}{A^5} ,
	\label{Densidade de Energia 1}
	\end{eqnarray}
and considering that  $\rho = \frac{3}{l^2A}$, we have
	\begin{eqnarray}
	\rho_{GUP} = \rho - \frac{1}{6}\pi\alpha l^2\rho ^2 + \frac{2}{27}\pi ^{2}\alpha ^2 l^4\rho ^3 
	+ \frac{1}{27}\pi^3\alpha\theta B l^6\rho ^4 + \frac{16}{405}\pi^4\theta^2 B^2l^8\rho ^5
	- \frac{16}{405}\pi^4\alpha ^2\theta Bl^8\rho ^5.
	\label{Densidade de Energia 2}
	\end{eqnarray}
	
{At this point, we use the laws of thermodynamics of black holes to determine the entropy of the BTZ black hole 
in a noncommutative background as follows}
\begin{eqnarray}
S_{GUP} &=&\int \frac{1}{T_{GUP}}\frac{\partial M}{\partial\hat{r}_+}d\hat{r}_{+}
=4\pi\int\left(1-\frac{\theta B}{4r^2_+}  \right)\left\{ \left(1+\frac{\theta B}{4{r}_{+}^2}\right)
\left[1 - \frac{\alpha}{4{r}_{+}}\left(1-\frac{\theta B}{4r_+^2} \right) + \frac{\alpha ^{2}}{8{r}_{+}^{2}}
\left(1-\frac{\theta B}{2r_+^2}\right)\right] \right\}dr_{+},
\nonumber\\
&=&4\pi r_{+} +\frac{\pi\theta^2B^2}{12r_+^3}-\pi\alpha\ln(r_+)-\frac{1}{8}\frac{\pi\alpha\theta B}{r_+^2}-\frac{1}{2}\frac{\pi\alpha^2}{r_+}
+\frac{1}{12}\frac{\pi\alpha^2\theta B}{r_+^3}+\cdots,
\end{eqnarray}
and that in terms of the entropy $ S=4\pi r_+=4\pi l\sqrt{M} $, we find
\begin{equation}
\label{Entgup}
S_{GUP} \leq S +\frac{16\pi^4\theta^2B^2}{3S^3}  -  \alpha\pi \ln\left({S}\right) 
- \frac{2\pi^3\alpha\theta B}{S^2}  -  \frac{2\pi ^2\alpha^2}{S} +\frac{16}{3}\frac{\pi^4\alpha ^2 \theta B}{S^3}
+\cdots .
\end{equation}
We have obtained corrections to the entropy through tunneling formalism using the Hamilton-Jacobi method 
due to the effects of GUP. Notice that from the above equation for $\alpha =0 $, we have that the noncommutative correction to the entropy occurs only at second order in the parameter $\theta$. 
Besides, we have obtained logarithmic corrections to the entropy of the BTZ black hole.
{Recently, in~\cite{Bargueno:2016qhu} the authors applied a method based on quantum-mechanical corrections to the Newtonian potential and calculated quantum-mechanical corrections to the entropy of a Schwarzschild black hole.	
They found a logarithmic correction of the type $ -\pi|\gamma|\ln(S_{BH}) $ where $ S_{BH}=A_{BH}/4 $ and $ A_{BH} $ is the area of the black hole horizon.  Comparing with our result in Eq. (\ref{Entgup}) we find that we have obtained a similar logarithmic correction given by $  -\alpha\pi \ln\left({S}\right) $. 
In Ref.~\cite{Das:2001ic} logarithmic corrections for entropy
of a BTZ black hole  were obtained employing a path integral approach for the gravitational partition function. 
The logarithmic correction term obtained by the authors is of the form $ -3/2\ln S 
 $. In our calculations the factor $ 3/2 $ is obtained if we fix $ \alpha=3/(2\pi) $. Furthermore,  by assuming $ \theta=0 $ in Eq. (\ref{Entgup}) our result agrees with that found in~\cite{Das:2001ic}. 
Moreover, it has also been investigated by the authors in~\cite{Bargueno:2015tea} semiclassical corrections for the entropy of a Schwarzschild black hole by applying the path integral method. The corrections obtained for the entropy due to the GUP become dominant for $\alpha$ being of the order of unity. Thus, the choice of $\alpha = 3/(2\pi)$ in our result is in accordance with the arguments of Ref.~\cite{Bargueno:2015tea}.
Therefore, a more detailed analysis of these results 
could be made 
by applying 
the approaches used by the authors in~\cite{Bargueno:2016qhu}, \cite{Das:2001ic}, \cite{Bargueno:2015tea},  
to the metric (\ref{metrica BTZ}) of a noncommutative BTZ black hole. We shall consider this study in a forthcoming publication.
}

\begin{figure}[htb]
\centering
{\includegraphics[scale=.9]{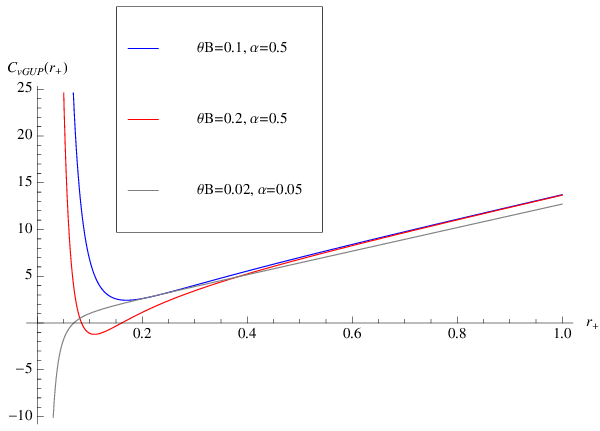}}
\caption{Specific heat capacity (Eq. (\ref{Cvgup})). Plot $ C_{vGUP} $ vs. $r_{+}$.  }
\label{capacity}
\end{figure}

\begin{figure}
\centering
{\includegraphics[scale=.9]{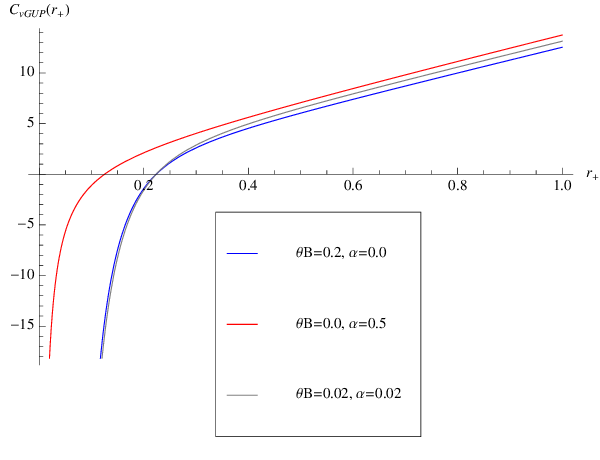}}
\caption{Specific heat capacity. From the top to bottom, the curves for $ (C_{v\theta}, C_{v\alpha GUP},C_{v\theta GUP}) $ vs. $r_{+}$.  }
\label{capacity2}
\end{figure}

The correction of the specific heat capacity at constant volume 
$C_v=T_h\left(\frac{\partial S}{\partial T_h} \right)_v=8\pi^2 l^2 T_{h}=4\pi r_{+}$, reads
\begin{eqnarray}
\label{Cvgup}
C_{vGUP} &=& 8\pi^2 l^2 T_{GUP}
= 8\pi^2 l^2\left[ \frac{r_+}{2\pi l^2} -\frac{\theta^2 B^2}{32\pi l^2 r_{+}^{3}}+\frac{\alpha}{8\pi l^2} 
-\frac{\alpha\theta B}{32\pi l^2 r_{+}^2} 
-\frac{\alpha^2}{16\pi l^2 r_+}+\frac{\alpha^2\theta B}{32\pi l^2 r_{+}^3}  +\cdots\right].
\end{eqnarray}
{Now if $ \alpha=0 $ in Eq. (\ref{Cvgup})  we have 
\begin{eqnarray}
\label{Cvtheta}
C_{v,\theta} &=& 4\pi r_{+}\left(1 -\frac{\theta^2 B^2}{16 r_{+}^{4}}+\cdots\right).
\end{eqnarray}
Note that the specific heat vanishes at the point $r_{+} ={\sqrt{\theta B}}/{2}$ and is positive for $r_{+} >{\sqrt{\theta B}}/{2}$. On the other hand if $ \theta B=0 $ in Eq. (\ref{Cvgup}) the $C_{GUP}$ is now given by
\begin{eqnarray}
\label{Cvalpha}
C_{v\alpha GUP} &=& 4\pi r_{+}\left(1+\frac{\alpha }{4r_{+}} 
-\frac{\alpha^2}{8 r_{+}^2} +\cdots\right),
\end{eqnarray}
which vanishes at the point $ r_{+} =\alpha/4$. In addition, neglecting the last two terms in (\ref{Cvgup}) and considering the case where $ \alpha = \theta B$ we obtain
\begin{eqnarray}
\label{Cvguptheta}
C_{v\theta GUP} &=& 
4\pi r_{+}\left(1+\frac{ \theta B}{4r_{+}} 
 -\frac{( \theta B)^2}{16 r_{+}^3} -\frac{( \theta B)^2}{16 r_{+}^{4}} +\cdots\right),
\end{eqnarray}
which is also zero at  $ r_{+} ={\sqrt{ \theta B}}/{2}$.
In terms of the $ T_h $ and $ C_v $ the Eq. (\ref{Cvgup}) becomes }
\begin{eqnarray}
C_{vGUP} &\leq &8\pi^2 l^2\left[ T_{h} -\frac{\theta^2 B^2}{256\pi^4l^8 T^3_{h}}+\frac{\alpha}{8\pi l^2} 
-\frac{\alpha\theta B}{128\pi^3 l^6 T_h^2} 
-\frac{\alpha^2}{32\pi^2 l^4 T_h}+\frac{\alpha^2\theta B}{256\pi^4 l^8 T_h^3}  +\cdots\right]
\nonumber\\
&=&C_v -\frac{2\pi^4\theta^2 B^2}{C^3_{v}}+\pi\alpha -\frac{\pi^3\alpha\theta B}{C_v^2} 
-\frac{\pi^2\alpha^2}{C_v}+\frac{2\pi^4\alpha^2\theta B}{C_v^3}+\cdots.
\end{eqnarray}	
Observe that Figs. \ref{capacity}  and \ref{capacity2} show the behavior of the specific heat capacity at constant volume.
{In Fig. \ref{capacity}, the graph shows that $C_{vGUP}$ is positive, e.g., for $\theta B=0.1$ and $\alpha=0.5$, indicating that the BTZ black hole 
in a noncommutative background is stable for this choice of parameters. For $\theta B=0.2$ and $\alpha=0.5$ one achieves two points where the specific heat vanishes, with an unphysical region in between. 
In the latter example, i.e. $\theta B=0.02$ and $\alpha=0.05$, one achieves one point where the specific heat capacity vanishes before entering into an unphysical zone. In the latter two cases the black hole decreases its size until achieve a critical radius where it ceases to evaporate and becomes a remnant. At these examples the last case has the lower critical radius.
In Fig. {\ref{capacity2}} was analyzed the behavior of the specific heat capacity for the cases, $\alpha=0$, $\theta B=0$ and $\theta B = \alpha$, respectively. All of them indicating presence of remnants.
Our result is similar to that found by the authors in Ref.~\cite{Rahaman:2013gw}.} 
{This can best be understood by considering the exact formula of temperature and specific heat. For the temperature we have the following exact expression
\begin{eqnarray}
\label{texact}
T_{GUP}=2\tilde{T}_{H}\Big(1+\frac{\alpha l_{p}}{\Delta x}\Big)^{-1}
\left[ 1+\sqrt{1-\frac{4}{(1+\frac{\Delta x}{\alpha l_p})^2}}\right]^{-1},
\end{eqnarray}
where $ \tilde{T}_{H} $ is the Hawking temperature obtained in Eq. (\ref{TH}).
Furthermore, an expression for the maximum temperature, $ T_{GUP}\leq T_{max}=\tilde{T}_{H} $
  can be obtained from Eq. (\ref{texact}) when 
$ \Delta x=\alpha l_p$ or $ (\hat{r}_{min}=\alpha l_p/2 )$, 
where the black hole reaches the minimum mass, i.e,  
\begin{eqnarray}
M_{min}=\frac{\alpha^2 l_p^2}{2 l^2}\Big(1-\frac{2\theta B}{\alpha^2 l_p^2} \Big). 
\end{eqnarray}
In addition, to compute the emission rate we use the Stefan-Boltzmann law in a three-dimensional spacetime~\cite{Tawfik:2013uza}
$
{dM}/{dt} \propto T_{GUP}^3.
$
Thus, for the emission rate we obtain 
\begin{eqnarray}
\frac{dM}{dt} \propto 8\tilde{T}^3_{H}\Big(1+\frac{\alpha l_p}{\Delta x}\Big)^{-3}
\left[ 1+\sqrt{1-\frac{4}{(1+\frac{\Delta x}{\alpha l_p})^2}}\right]^{-3}.
\end{eqnarray}
Now we make the scaling $\Delta x=\alpha l_p f(M)$ into Eq.~(\ref{texact}) to obtain the following exact expression 
for the specific heat
\begin{eqnarray}\label{cvg-exact}
C_{GUP}&=&\Big( \frac{dT_{GUP}}{dM} \Big)^{-1}
\nonumber\\
&=&\frac{2 \pi \alpha l_ p l^2 \Big(f(M)+1\Big)^3} {f^{\prime}(M) D }\sqrt{1-\frac{4}{\Big(f(M)+1\Big)^2}} \left[\sqrt{1-\frac{4}{\Big(f(M)+1\Big)^2}}+1\right]^2,
\end{eqnarray}
where $ f(M) $ is a function of the mass $ M $ with $ f^{\prime}(M)\neq 0 $ and 
\begin{eqnarray}
D=\left[\alpha^2 l^2_p f(M)\Big(f^2(M)+3 f(M) + 2\Big) +\theta B \Big(1+f(M)\Big)\right]\left[\sqrt{1-\frac{4}{\Big(f(M)+1\Big)^2}}+1\right] 
-8 \alpha^2 l^2_pf(M) . 
\end{eqnarray}
Note that for the minimum mass $ \Delta x \rightarrow \alpha l_p $ we will have $ f(M)\rightarrow 1 $, so the specific heat (\ref{cvg-exact}) tends to zero 
(i.e., $C_{GUP}\rightarrow 0$). 
Therefore, this result indicates 
the black hole ceases to evaporate completely and becomes a remnant, which may help to enhance the information loss paradox discussions.
}

\newpage

\section{CONCLUSIONS}	\label{conclu}
In summary, using the Hamilton-Jacobi version of the tunneling formalism we have considered the metric of a noncommutative BTZ black hole and we have shown that the noncommutative corrections for Hawking temperature and entropy occur only in second order in the parameter $ \theta $. 
In addition, we have also shown that the product of entropy is dependent on the mass parameter $ M $ and becomes independent of the mass when $ \theta =0$.
Besides, by considering the GUP via the Hamilton-Jacobi method to calculate the imaginary part of the action, 
we have obtained quantum corrections for Hawking temperature, entropy and specific heat capacity of a noncommutative BTZ black hole. Moreover, in our calculations the GUP was introduced by the correction to the energy of a particle
due to gravity near the horizon. Thus, in our model the GUP allows us to find logarithmic corrections to the area law. {We have found that noncommutative BTZ black hole is stable for $r_+ > r_{\theta}={\sqrt{\theta B}}/{2}$. Furthermore, we also have analyzed the nature of the specific heat capacity and we have observed from Fig. \ref{capacity} and Fig. \ref{capacity2} the signal of presence of remnants as final stage of noncommutative BTZ black hole.}

\acknowledgments
We would like to thank CAPES and CNPq for financial support.

\end{document}